\documentclass[sigconf]{acmart}

\AtBeginDocument{%
  }

\usepackage{algorithm}
\usepackage{algorithmic}
\usepackage{amsmath}
\usepackage{bbding}
\usepackage{tcolorbox}
\usepackage{lipsum}
\usepackage{listings}
\usepackage{multirow}
\usepackage{graphicx}
\usepackage{cleveref}
\usepackage{hyperref}
\usepackage{subcaption}
\usepackage{pifont}
\usepackage{fontawesome}
\setcopyright{acmlicensed}
\copyrightyear{2018}
\acmYear{2018}
\acmDOI{XXXXXXX.XXXXXXX}
\acmConference[Conference acronym 'XX]{Make sure to enter the correct
  conference title from your rights confirmation emai}{June 03--05,
  2018}{Woodstock, NY}
\acmISBN{978-1-4503-XXXX-X/18/06}

\begin{document}

\title{From Prompting to Alignment: A Generative Framework for Query Recommendation}

\author{Erxue Min$^{\ast}$}
\email{erxue.min@gmail.com}
\affiliation{
  \institution{Baidu Inc.}
  \city{Beijing}
  \country{China}
}

\author{Hsiu-Yuan Huang$^{\ast}$}
\affiliation{
  \institution{National Key Laboratory for Multimedia Information Processing, Peking University}
  \city{Beijing}
  \country{China}}
\email{huang.hsiuyuan@stu.pku.edu.cn}

\author{Xihong Yang$^{\ast}$}
\email{xihong\_edu@163.com}
\author{Min Yang}
\email{yangmin11bit@gmail.com}

\author{Xin Jia$^{\dagger}$}
\email{jemmryx@pku.edu.cn}

\affiliation{%
  \institution{Baidu Inc.}
  \city{Beijing}
  \country{China}
}

\author{Yunfang Wu}
\affiliation{
  \institution{National Key Laboratory for Multimedia Information Processing, Peking University}
  \city{Beijing}
  \country{China}}
\email{wuyf@pku.edu.cn}

\author{Hengyi Cai}
\affiliation{
  \institution{Chinese Academy of Sciences}
  \city{Beijing}
  \country{China}}
\email{caihengyi@ict.ac.cn}

\author{Junfeng Wang}
\email{wangjunfeng@baidu.com}

\author{Shuaiqiang Wang}
\email{shqiang.wang@gmail.com}

\author{Dawei Yin}
\email{yindawei@acm.org}
\affiliation{
  \institution{Baidu Inc.}
  \city{Beijing}
  \country{China}
}

\renewcommand{\shortauthors}{Trovato et al.}
\newcommand{\todo}[1]{\textcolor{red}{ (#1)}}
\newcommand{\added}[1]{\textcolor{blue}{ (#1)}}

\begin{abstract}
In modern search systems, search engines often suggest relevant queries to users through various panels or components, helping refine their information needs. Traditionally, these recommendations heavily rely on historical search logs to build models, which suffer from cold-start or long-tail issues. Furthermore, tasks such as query suggestion, completion or clarification are studied separately by specific design, which lacks generalizability and hinders adaptation to novel applications. 
Despite recent 
attempts to explore the use of LLMs for query recommendation, these methods mainly rely on the inherent knowledge of LLMs or external sources like few-shot examples, retrieved documents, or knowledge bases, neglecting the importance of the calibration and alignment with user feedback, thus limiting their practical utility. To address these challenges, we first propose a general Generative Query Recommendation (GQR) framework that aligns LLM-based query generation with user preference. Specifically, we unify diverse query recommendation tasks by a universal prompt framework, leveraging the instruct-following capability of LLMs for effective generation. Secondly, we align LLMs with user feedback via presenting a CTR-alignment framework, which involves training a query-wise CTR predictor as a process reward model and employing list-wise preference alignment to maximize the click probability of the generated query list. Furthermore, recognizing the inconsistency between LLM knowledge and proactive search intents arising from the separation of user-initiated queries from models, we align LLMs with user initiative via retrieving co-occurrence queries as side information when historical logs are available. We validate the effectiveness of our approach across three distinct scenarios within Baidu's conversational search services, demonstrating remarkable improvements in both CTR (60\%+) and user experience.
\end{abstract}

\ccsdesc[500]{Information systems~Query suggestion}
\ccsdesc[500]{Computing methodologies~Natural language generation}

\keywords{Large Language Model, Query Recommendation, Preference Alignment}

\received{20 February 2007}
\received[revised]{12 March 2009}
\received[accepted]{5 June 2009}
\thanks{$^{\ast}$: Equal contribution. $^{\dagger}$: Corresponding author.}
\maketitle

\begin{figure}[!t]
  \centering
\includegraphics[width=\linewidth]{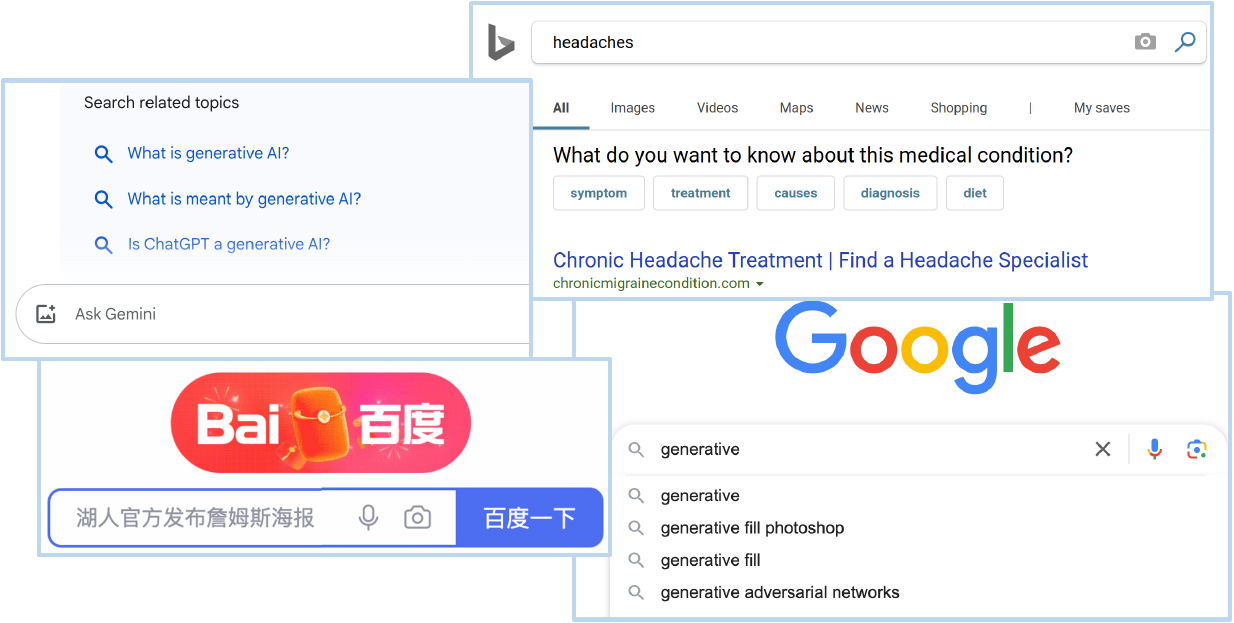}
  \caption {Examples of query recommendations in web search engines and conversational search systems. Clickable queries are displayed in various forms to increase search efficiency.\looseness=-1}
  \label{fig:top}
\end{figure}

\section{Introduction}
Search engines often employ various query recommendation modules to assist users in producing effective queries to meet their information needs, as demonstrated in Figure~\ref{fig:top}. 
For example, query completion \cite{2014QueryAutoCompletion,2015QueryAutoCompletion,SurveyofQueryAutoCompletion,ZamaniGeneratingClarifyingQuestions}
offers real-time query predictions or completions based on the user’s input prefix to reduce spelling mistakes, discover relevant search terms, or increase search efficiency.
Query suggestion \cite{uddin2018multitask,2019ClickFeedbackAwareQueryRecommendation,mustar2020usingBERTandBART,chen2020incorporatingbehavioralhypothesesquery,lee2024enhancedfacetgenerationllm,MiningExploratoryQueries_Dou} presents users with relevant queries based on popular search trends, user behavior, and historical data, helping to explore different search intents.
Query clarification \cite{ZamaniUserInteractions,2020clarifyingQuestion,2021FacetDrivenGenerationClarifyingQuestionsSearch,2022GeneratingClarifyingQuestionsWeb,ZamaniGeneratingClarifyingQuestions,pyatkin-etal-2023-clarifydelphi} 
addresses ambiguous queries by offering clarifying questions or displaying several candidate query facets to help refine users' search intent. 
We collectively term these tasks as ``Query Recommendation'' due to their shared objective of presenting clickable query candidates.

Conventional query recommendation (CQR) systems span from statistical approaches \cite{10.1145/1963405.1963424,10.1145/2505515.2505661,6816668,10.1145/3020165.3022129,10.1145/3077136.3080652} to neural network methods \cite{10.1145/2806416.2806493,10.1145/3132847.3133010,10.1145/3397271.3401331,ZamaniConversationsClarifyingQuestions,2017NeuralLanguageModel4QAC,jaech2018personalizedlanguagemodelquery}. Although these methods have achieved remarkable commercial success, three fundamental challenges persist in real-world applications:
\textbf{(C1) Sparsity in Historical Interactions.} Current CQR systems \cite{10.1145/1963405.1963424,10.1145/2505515.2505661,2017NeuralLanguageModel4QAC,jaech2018personalizedlanguagemodelquery} 
rely heavily on query log co-occurrence patterns, making them ineffective for cold-start scenarios and long-tail queries with insufficient interaction data. 
\textbf{(C2) Cross-Scenario Generalization Gap.} Existing solutions \cite{2022LearningMultipleIntentforSearchQueries,RevisitingQueryFacetGeneration,StochasticOptimization_MultipleQueryIntent,2023QuerySubIntentMining,Dou2023ImprovingSearchClarificationSERP} 
employ scenario-specific designs respectively, increasing the development overhead and limiting cross-task knowledge transfer. Furthermore, it hinders adaptation to emerging applications like conversational search systems that demand contextual understanding.
\textbf{(C3) User Preference Misalignment.} While recent attempts have leveraged Large Languages (LLM) \cite{MiningExploratoryQueries_Dou,MultiTurnClarificationDou,ZeroShot-CQGen-4ConversationalSearch,2024MultimodalQuerySuggestion} or Retrieval Augmented Generation (RAG) \cite{bacciu2024generatingqueryrecommendationsllms,shen2024enhancing,wang2024richragcraftingrichresponses} to mitigate the aforementioned limitations, 
they fail to align generated queries with actual user preferences. The absence of feedback calibration often results in semantically plausible but practically ineffective recommendations.

To overcome these challenges, we propose \textbf{Generative Query Recommendation} (GQR), a novel paradigm that formulates the query recommendation tasks as a generation process aligned with user preferences.
To tackle challenge \textbf{C1}, we utilize LLMs as the backbone of our framework, harnessing their inherent world knowledge and strong semantic comprehension to effectively handle cold-start scenarios and long-tail queries without requiring heavily on past interactions. To tackle challenge \textbf{C2}, we develop a universal prompt template adaptable to various query-centric scenarios, from general web search to context-sensitive conversational systems. This framework supports flexible output formatting and integration of contextual signals like conversation history or retrieved documents.
To tackle challenge \textbf{C3}, We align LLMs with user preference from two aspects: \textit{(a) User Feedback}. We propose a CTR-based alignment framework, which treats each query generation as a thinking step, and trains a CTR predictor as a Process Reward Model (PRM) \cite{lightman2023let} based on user click signals to optimize the generation quality. \textit{(b) User Initiative}. When search logs are available, we augment LLM inputs with retrieved co-occurrence queries to bridge the gap between model generation and proactive search patterns.

To the best of our knowledge, this is the first generative framework for general query-centric tasks in commercial search systems. We anticipate this paper to provide our practical experiences and novel insights for incorporating LLMs into the query recommendation. The main contributions are outlined as follows:
\begin{itemize}
    \item \textbf{Unified Task Design}. We unify diverse query-centric tasks into a Generative Query Recommendation framework using a general prompt template, applicable to various intricate or contextually relevant settings.
    \item \textbf{User Feedback Alignment}. We propose a CTR-alignment framework to maximize the users' click preferences. As far as we know, we are the first to align LLM generation with click feedback in large-scale commercial search systems, which is non-trivial due to the noisy and imbalanced nature of click behaviors.
    \item \textbf{User Initiative Alignment}. We further devise an initiative-alignment strategy, which introduces co-occurrence user-initiated queries as external references, augmenting LLMs with proactive search preference.
    \item \textbf{System Deployment}. We design an effective system workflow to continuously improve the overall performance, which comprises a periodic update strategy and multiple exploration strategies during both training and inference.
    \item \textbf{Evaluation}. We conduct extensive offline and online experiments to validate the effectiveness of the GQR system. The results show that the deployment of the system can significantly improve the usability and applicability of various query components, achieving up to 60\%+ CTR improvements compared with LLM baselines.
\end{itemize}


\begin{table}[]
\centering
\caption{Comparison of CQR, existing LLM-based methods and our GQR paradigm, in terms of 1) Effectiveness with sparse interactions, 2) Generalization ability to varied domains and 3) Alignment with user preference.}
\scalebox{0.9}{
\begin{tabular}{c|c|c|c}
\hline
    & Sparsity & Generalization & User Alignment \\ \hline
CQR & \faThumbsDown      & \faThumbsDown            & \faThumbsOUp            \\ \hline
LLM & \faThumbsOUp      & \faThumbsOUp            & \faThumbsDown            \\ \hline
GQR & \faThumbsOUp      & \faThumbsOUp            & \faThumbsOUp            \\ \hline
\end{tabular}
}
\end{table}

\section{Related Work}
Our work is primarily related to three research directions: \textbf{(1)} Query Prediction. Our approach introduces a general generative framework for diverse query-centric tasks
including query auto-completion \cite{2012AQC10.1145/2348283.2348364, 2014QAC10.1145/2566486.2568009, 2016QAC10.1145/2911451.2914686,2017NeuralLanguageModel4QAC,jaech2018personalizedlanguagemodelquery,wang2018realtime}, query suggestion \cite{10.1145/3397271.3401331,chen2020incorporatingbehavioralhypothesesquery,lee2024enhancedfacetgenerationllm,MiningExploratoryQueries_Dou,2024MultimodalQuerySuggestion} and query clarification \cite{ZamaniGeneratingClarifyingQuestions,MultiTurnClarificationDou,ZeroShot-CQGen-4ConversationalSearch,FindingDimensions4Queries,2014ExtendingFacetedSearchtoWeb,201310.1145/2484028.2484097}, etc. 
\textbf{(2)} LLM-based Recommendation. LLMs have been extensively studied to enhance various aspects of recommender systems using effective techniques such as in-context learning \cite{li2023gpt4recgenerativeframeworkpersonalized,sun2024snippetbasedconversationalrecommender,bacciu2024generatingqueryrecommendationsllms}, fine-tuning \cite{qiu2021u,wu2021userbertcontrastiveusermodel,10.1145/3488560.3498495,10.1145/3485447.3511977,zhang2023bridginginformationgapdomainspecific,zhang2024collmintegratingcollaborativeembeddings}, and prompt tuning \cite{2022TowardsUnifiedCRSviaKnowledgeEnhancedPromptLearning, cui2022m6recgenerativepretrainedlanguage,2023TALLRec,10.1007/978-3-031-56063-7_42,yang2023palrpersonalizationawarellms}, etc. These approaches primarily focus on improving item-user correlations, while our GQR paradigm fundamentally differs by generating textual queries optimized for high user click preference.
\textbf{(3)} Preference Alignment for LLM.  
The effectiveness and applicability of LLMs largely depend on aligning models with human intentions by preference learning \cite{schulman2017PPO,NEURIPS2023_DPO,pmlr-v238-gheshlaghi-azar24a,meng2024simposimplepreferenceoptimization,hong-etal-2024-orpo,ethayarajh2024_KTO}. Although there have been some studies on learning with noisy preference labels \cite{chowdhury2024provablyrobustdpoaligning,chen2024entityalignmentnoisyannotations}, alignment with click preference of large-scale users in commercial systems remains uncharted territory. Due to page constraints, we include a detailed introduction of the related work in Appendix \ref{sec:RelatedWork}.

\section{Methodology}
In this section, we elaborate on our GQR framework in detail. We first propose a general prompt framework for query recommendation tasks. Subsequently, we present our method for CTR alignment, followed by a strategy for aligning users' initiative search intent. Furthermore, we introduce our strategies for enterprise deployment regarding periodic updates and exploration.

\subsection{A Unified Prompt Framework for Query Recommendation Tasks}
\label{sec:promptframe}
In this section, we propose a universal prompt template for diverse query-centric tasks, which frames all sorts of query prediction or recommendation tasks as a generation process. GQR adopts a pre-trained LLM as its foundation, leveraging its strong instruct-following capabilities and inherent world knowledge to generate valuable queries that meet different requirements, such as cold-start, long-tail, complicated, or contextually relevant scenarios.


To elaborate, as illustrated in  prompt template \ref{box:prompt}, given a query input text $q$ and a set of (optional) side information $S=\{s_1,s_2,\ldots,s_k\}$, a pre-trained LLM $\mathcal{M}$ is instructed to generate a list of recommended queries $RQ=\{q_1,q_2,\ldots,q_N\}$ along with a collection of (optional) auxiliary texts $T=\{t_1,t_2,\ldots,t_l\}$.
The side information $S$ can encompass any task-specific auxiliary information. For instance, in a general search scenario, $S$ may comprise past queries from the ongoing session, while in a conversational search context, $S$ can be previous queries and the AI system response for the current query.
Likewise, the auxiliary texts $T$ are task-dependent. For instance, in a choice-based query clarification task of an AI assistant, it can be a guiding sentence like ``What do you want to know?'' or a prefix/suffix text aiding in forming a complete query by concatenating with the selected short option. We require LLMs to generate multiple queries in a single prompt call for two reasons: (1) Cost control, considering the high computational cost of invoking LLMs in online systems; (2) It allows us to instruct LLMs to generate queries satisfying specific interrelationships, such as semantic exclusivity.

An effective method to instruct LLM $\mathcal{M}$ to generate responses in the desired format is through In-Context Learning (ICL) \cite{brown2020language},
which places high demands on the capabilities of the model and the quality of demonstration examples. Therefore, for real-world deployments considering cost and latency constraints, a more practical approach is to fine-tune a relatively smaller LLM (e.g., 7B, 13B) tailored for the specific task. Concretely, with a GQR training dataset $\mathcal{D}_{sft}=\{(\mathbf{x}^i, \mathbf{y}^i)\}_{i=1}^N$ annotated by expert or large LLMs like GPT, where $\textbf{x}^i$ and $\textbf{y}^i$ represent the $i$-th prompt and response correspondingly, SFT can be executed by minimizing the  cross-entropy loss as follows:
\begin{equation}
\mathcal{L}(\theta)=-\frac{1}{N} \sum_{i=1}^N \sum_{t=1}^T \log \mathbb{P}_\theta\left(y_t^i \mid \mathbf{x}^i, \mathbf{y}_{1: t-1}^i\right)
\end{equation}

\subsection{User Feedback Alignment}
\label{sec:ctr_align}
Once we have a customized LLM model for a specific query recommendation task and integrate it into an online service, we start gathering extensive impression click\footnote{In addition to impression clicks, other user feedback signals such as dwell time, scroll depth, and conversion events (e.g., sign-ups or purchases) may also be available in some systems. However, in this paper, we focus solely on impression click signals, given their most widespread availability.} logs from real users. Typically, the initial LLM falls short of perfection due to challenges in accurately capturing users' actual interests or preferences. Consequently, aligning LLMs with user click preference is crucial to improve user engagement and enhance user experiences.
However, due to the inherent randomness, noise and significant imbalance in user click behaviours, constructing preference labels for each response (i.e., a query list) or developing reward models directly poses non-trivial challenges. Therefore, in our work, we attempt to estimate the CTR of each Recommended Query (RQ) and align the LLM responses to high-CTR ones.
Formally, when presented with a prompt $\textbf{x}_i$ containing the target user query and side information, the LLM generates an output response $\mathbf{y}$ comprising auxiliary texts and a list of RQs, denoted as 
$[T|q_1|q2|\ldots|q_N]$.


\definecolor{colframecolor}{RGB}{55,98,175}
\definecolor{colbackcolor}{RGB}{218,227,243}

\newtcolorbox[auto counter]{mybox}[2][]{
colframe=colframecolor,
colback=colbackcolor, 
coltitle=white, 
fonttitle=\bfseries, 
title=\thetcbcounter. {#2},
#1
}

\begin{figure}[t]
\begin{mybox}[label=box:prompt]{Prompt Template for Query Recommendation}
  \textbf{Instructions:} \\
  You are an expert in query recommendation, you will be given a search query $q_{user}$ along with a set of side information $S$, please generate $N$ relevant queries that users are most likely to click, and a set of auxiliary texts $T$.\\
  \textbf{Inputs:}\\
  <$q_{user}$> //search query (required)\\
  <$s_1$> //side information 1 (optional)\\
  $\ldots$\\
<$s_k$> //side information k (optional)
  \tcblower
  \textbf{Response:} \\
  <$t_1$> //auxiliary text 1 (optional)\\
  $\ldots$\\
<$t_l$> //auxiliary text $l$ (optional)\\
  <$q_1$><SEP><$q_2$><SEP>\ldots<SEP><$q_N$> //RQs (required)
\end{mybox}
\end{figure}

The primary aim is to maximize the expected probability of these queries being clicked\footnote{Click probability is not equivalent to CTR in some case, we omit this difference and interchangeably use the two terms for simplicity.}.
The optimization objective is defined as:
\begin{equation}
\max _{\pi_\phi} \mathbb{E}_{\mathrm{x} \sim D, \mathrm{y} \sim \pi_\phi(\mathrm{y} \mid \mathrm{x})}[r_\theta(\mathrm{x}, \mathrm{y})-\beta R_\phi(\mathbf{x},\mathbf{y})].
\end{equation}
where $R_\phi(\mathbf{x},\mathbf{y})$ denotes the regularization term, which can be a weighted sum of pretraining loss and KL loss between policy model $\pi^{RL}$ and SFT model $\pi^{SFT}_\phi$ in PPO \cite{schulman2017PPO}.
The reward function $r_\theta(\mathbf{x}, \mathbf{y})$ is defined as the likelihood of the recommended queries in $RQ$ being clicked. The calculation formula of $r_\theta(\mathbf{x}, \mathbf{y})$ is dependent on the characteristics of the component displaying RQs. Assuming that each click on an exposed RQ is an independent event, we present the equations for two component types:

For multi-choice components:
\begin{equation}
\label{eq:multi}
\begin{split}
r_\theta(\mathbf{x}, \mathbf{y})_{multi} =& \mathbb{P}(has\_click|q_{user},q_1,\ldots,q_N)\\
=& 1-\mathbb{P}(no\_click|q_{user},q_1,\ldots,q_N)\\
=&1-\prod_{k=1}^N(1-\mathbb{P}(c_k|q_{user},q_k,C_k)),\\
\end{split}  
\end{equation}
and for single-choice components where each click is exclusive:
\begin{equation}
\label{eq:single}
\begin{split}
r_\theta(\mathbf{x}, \mathbf{y})_{single} = &\sum_{k=1}^N \mathbb{P}(c_k|q_{user},q_k,C_k)).
\end{split}
\end{equation}
Here, $c_k$ represents the event that $q_k$ is clicked, and $C_k$ represents the context in which $q_k$ is displayed, such as nearby exposed RQs that might influence the CTR of $q_k$ or other information like AI responses.
Consequently, the process of aligning with user click preference involves two primary subtasks:  1) Developing a query-wise CTR predictor and 2) List-wise CTR alignment.

\subsubsection{Query-wise CTR predictor}

\begin{figure}[ht]
  \centering
  \includegraphics[width=\linewidth]{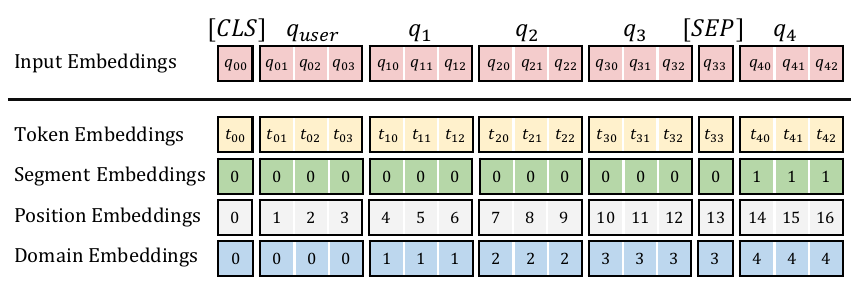}
  \caption {Input representation of the CTR predictor. The input embeddings are the sum of the token embeddings, the segmentation embeddings, the position embedding and the domain embeddings.}
  \label{fig:ctr_model}
\end{figure}

To obtain the click probability $p_k=\mathbb{P}(c_k|q_{user},q_k,C_k))$ of each query $q_k$, we need to train a CTR predictor. 
To simplify, we assume that the CTR of $q_k$ is only primarily influenced by preceding generated queries, i.e., $p_k=\mathbb{P}(c_k|q_{user},q_1,\ldots,q_{k-1},q_{k}))$.
Given the noise and imbalance present in impression click logs, along with the substantial amount of data (ranging from millions to billions) required to train an effective CTR predictor in practical applications, there is no benefit in utilizing large LLMs over small models.
Therefore, we employ the BERT \cite{devlin2018bert} architecture as the backbone of our CTR predictor.
As depicted in Figure~\ref{fig:ctr_model}, we input the user query $q_{user}$, preceding recommended queries $q_1,\ldots,q_{k-1}$, and the target recommended query  $q_k$ to predict, separated by a \textbf{[SEP]} token and initiated with a \textbf{[CLS]} token:
$$
X_{ctr}= \textbf{[CLS]}q_{user}q_1\ldots q_{k-1}\textbf{[SEP]}q_k.
$$
For each token $x_i$ in the input text, it can be represented as a sum of four distinct embeddings as follows:
$$
\mathbf{e}_i = \mathbf{e}_i^t + \mathbf{e}_i^p + \mathbf{e}_i^s + \mathbf{e}_i^d,
$$
where $\mathbf{e}_i^t$ denotes the token embedding, $\mathbf{e}_i^p$ is the position embedding, $\mathbf{e}_i^s$ indicates the segment embedding, $\mathbf{e}_i^d$ stands for the domain embedding, allowing the model to distinguish different parts of the input (e.g., different queries).
Given the embedding sequence $\mathbf{E}_{ctr}=[\mathbf{e}_1,\mathbf{e}_2,\ldots,\mathbf{e}_L]$, we leverage a BERT \cite{devlin2018bert} architecture to generate contextualized representations for each token:
$$
\mathbf{E}^c_{ctr} = [\mathbf{e}_1^c,\mathbf{e}_2^c,\ldots,\mathbf{e}_L^c] = \text{BERT}(\mathbf{E}_{ctr}),
$$
Subsequently, a two-layer Multilayer Perceptron (MLP) module operates on $\mathbf{e}_1^c$, which is the $\mathbf{CLS}$ token, to produce the logit, followed by a Sigmoid function to derive the final click probability score $p_k$:
$$
p_k = \text{Sigmoid}(\text{MLP}(\mathbf{e}_1^c))
$$
We construct a dataset for training the CTR predictor by leveraging impression click logs from our online conversational search system. Each log entry is structured as:
$$
[\mathbf{x}, T|q1|q2|\ldots|q_N, \mathbf{b}],
$$
where $\mathbf{x}$ is the prompt, $\textbf{b}\in \mathbb{R}^N$ is a one-hot or zero vector, with $\textbf{b}_k=1$ indicating that a user clicked on the displayed query $q_k$.
We extract $N$ training instances from each user log and construct a dataset $\mathcal{D}_{ctr}$ from logs spanning the last 14 days.
Consequently, the training objective of the CTR predictor is formulated as:
\begin{equation}
L(y, \hat{p}_\theta) = -\frac{1}{|\mathcal{D}_{ctr}|} \sum_{(x, y) \in \mathcal{D}_{ctr}} [ y \log(\hat{p}_\theta) + (1 - y) \log(1 - \hat{p}_\theta)]
\end{equation}
We optimize the loss function via Adam\cite{kingma2017adammethodstochasticoptimization} and obtain $\hat{p}_k$ as the estimated click probability of each query $q_k$.

\begin{algorithm}[!t]
\caption{\textbf{Iterative DPO Training for CTR Alignment}}
\label{iterDPO}
\flushleft{\textbf{Input}: initialized LLM $\pi_{\phi_0}^{DPO}$; trained CTR model $f_\theta(\cdot)$;  test data $X_{\text{test}}$; threshold $\delta$; beam size $B$} \\
\flushleft{\textbf{Initialize:}: $S_{\text{prev}} \gets 0, \pi_{\phi_0}^{DPO} \gets \pi_{\phi}^{SFT}, t \gets 1$} 
\vspace{-10pt}
\begin{algorithmic}[1]
\FOR {$\Delta S < \delta$}
\FOR{\underline{each input $x \in X_{train}$}}
\STATE {// Generate chosen response using beam search}
\STATE Initialize beam $\mathcal{B} \gets \{\}$;
\FOR{\underline{$k = 1$ to $N$}}
\STATE Initialize beam $\mathcal{B}_{\text{new}} \gets \emptyset$;4
\FOR{\underline{each partial sequence $q_{1:k-1} \in \mathcal{B}$}}
\STATE Generate queries $C$ from $\pi_{\phi_{t-1}}^{DPO}$ given $x$ and $q_{1:k-1}$;
\FOR{\underline{each $q_k \in C$}}
\STATE Form sequence $q_{1:k} = q_{1:k-1} \cup \{q_k\}$\;
\STATE Update  CTR using Equation \ref{eq:multi} or \ref{eq:single}\;
\STATE Add $(q_{1:k}, P)$ to $\mathcal{B}_{\text{new}}$\;
\ENDFOR
\ENDFOR
\STATE Select top $B$ sequences from $\mathcal{B}_{\text{new}}$ based on $P$\;
\STATE Update beam $\mathcal{B} \gets$ selected sequences\;
\ENDFOR
\STATE $Y_{\text{chosen}} \gets \mathcal{B}$\;
\STATE  {// Generate rejected response}
\FOR{\underline{$i = 1$ to $N^\text{budget}_r$}}
\STATE Randomly generate response $y_i$ from $\pi_{\phi_{t-1}}^{DPO}$ given $x$\;
\STATE Compute CTR using Equation \ref{eq:multi} or \ref{eq:single}\;
\ENDFOR
\STATE $Y_{\text{rejected}} \gets N_r$ sequences with lowest $P_i$ \;
\STATE Filter overlapping responses in $Y_\text{chosen}$\;
\STATE Get triplet $(\mathbf{x},\mathbf{y}_c,\mathbf{y}_r)$ using Equation (\ref{min_length})\;
\STATE $\mathcal{D}_t = \mathcal{D}_t \cup (\mathbf{x},\mathbf{y}_c,\mathbf{y}_r)$
\ENDFOR
\STATE {// DPO training and evaluation}
\STATE $\pi_{\phi_t}^{DPO}$ = \text{DPO}($\pi_{\phi_{t-1}}^{DPO}, X_\text{train}, Y_c, Y_r$)\;
\STATE Evaluate average score $S$ on $X_{\text{test}}$ using $\pi_{\phi_t}^{DPO}$ and $f_\theta(\cdot)$\;
\STATE $\Delta S \gets S - S_{\text{prev}}$, $S_{\text{prev}} \gets S, t \gets t+1$\; 
\ENDFOR
\flushleft{\textbf{return} {$\pi_{\phi_t}^{DPO}$}} 
\end{algorithmic}
\end{algorithm}

\subsubsection{List-wise Preference Alignment}
\label{sec:alignment}
Since we have obtained the CTR predictor as a PRM, we can align preferences for LLM $\pi^{sft}_\phi$ using RL frameworks like PPO \cite{schulman2017PPO}, DPO \cite{NEURIPS2023_DPO} or KTO \cite{ethayarajh2024_KTO}.
Given the challenges of hyper-parameter selection and code-level optimization of PPO\cite{schulman2017PPO}, we opt for iterative DPO\cite{NEURIPS2023_DPO} as our training framework.
The key concern of DPO lies in constructing the pair of chosen and rejected responses for each prompt. 
For the chose response, we aim to maximize $r_\theta(\mathbf{x},\mathbf{y}_{chosen})$, while rejected responses should exhibit a low CTR.

As demonstrated in Algorithm \ref{iterDPO}, in each epoch, beam search is utilized to produce $N_c$ candidate chosen responses with the highest cumulative CTR score for each input $x$. Subsequently, $N_r$ candidate responses with the lowest scores are selected from multiple random inferences.
Our empirical investigations unveiled two notable drawbacks associated with the CTR-maximum strategy:
\begin{itemize}
    \item \textbf{Overlapping queries:} The selection strategy based on CTR often leads to semantically similar or even identical queries in responses due to the similar patterns present in high CTR queries. To mitigate this issue, we leverage larger language models such as GPT to identify and filter out these redundancies.
    \item \textbf{Lengthy responses:} The CTR reward model tends to assign higher scores to longer queries, which is consistent with existing studies \cite{ouyang2022training,ye2024justice,saito2023verbosity}. This propensity can aggressively increase the length of LLM responses following multiple DPO training iterations. To counteract this effect, we opt to choose the response $\mathbf{y}_{i^*}^c$ and reject query $\mathbf{y}_{j^*}^r$ with the most similar lengths from the two sets, where
\begin{equation}
\label{min_length}
(i^*, j^*) = \arg \min_{i, j} | \text{len}(\mathbf{y}_i^c) - \text{len}(\mathbf{y}_j^r) |
\end{equation}
\end{itemize}

Consequently, for each input x, we generate a training instance represented as a $(\mathbf{x}, \mathbf{y}_i^c, \mathbf{y}_i^r)$ triplet.
During each iteration, we conduct DPO training on the produced dataset $\mathcal{D}_k$ to obtain LLM $\pi_{\phi_k}^{DPO}$ (DPO is detailed in Appendix \ref{sec:DPO} due to the limit of pages). Subsequently, this model is utilized to produce a response set $Y_k^{test}$ from a predetermined testing set $X_{test}$, and the average CTR score $S_k$ is determined using the CTR predictor.
The iterative process continues until the difference between $S_k$ and $S_{k-1}$ is less than a predefined threshold value $\delta$.

\begin{figure}[t]
    \centering
    \includegraphics[width=0.95\linewidth]{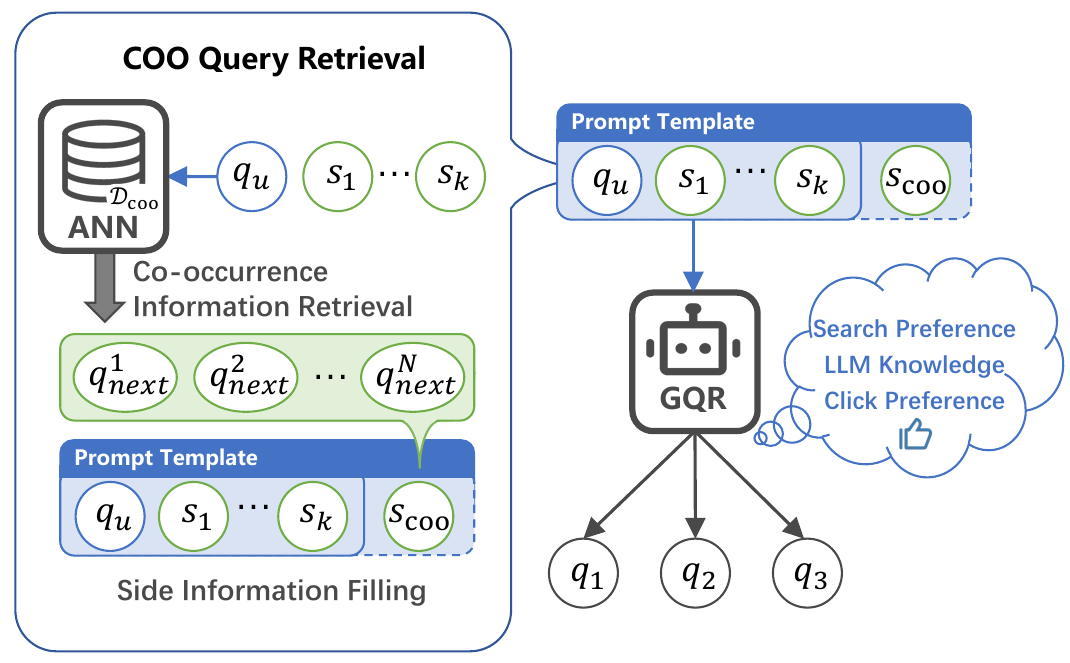} %
        \vspace{-1em}
    \caption{Illustration of the generation after introducing the co-occurrence retrieval stage. 
    The queries are generated by an implicit composition of LLM inherent knowledge, user click preference and co-occurrence information.
    }
    \label{fig:psp}
    \vspace{-1em}
\end{figure}

\begin{figure}[t]
    \centering
    \includegraphics[width=0.9\linewidth]{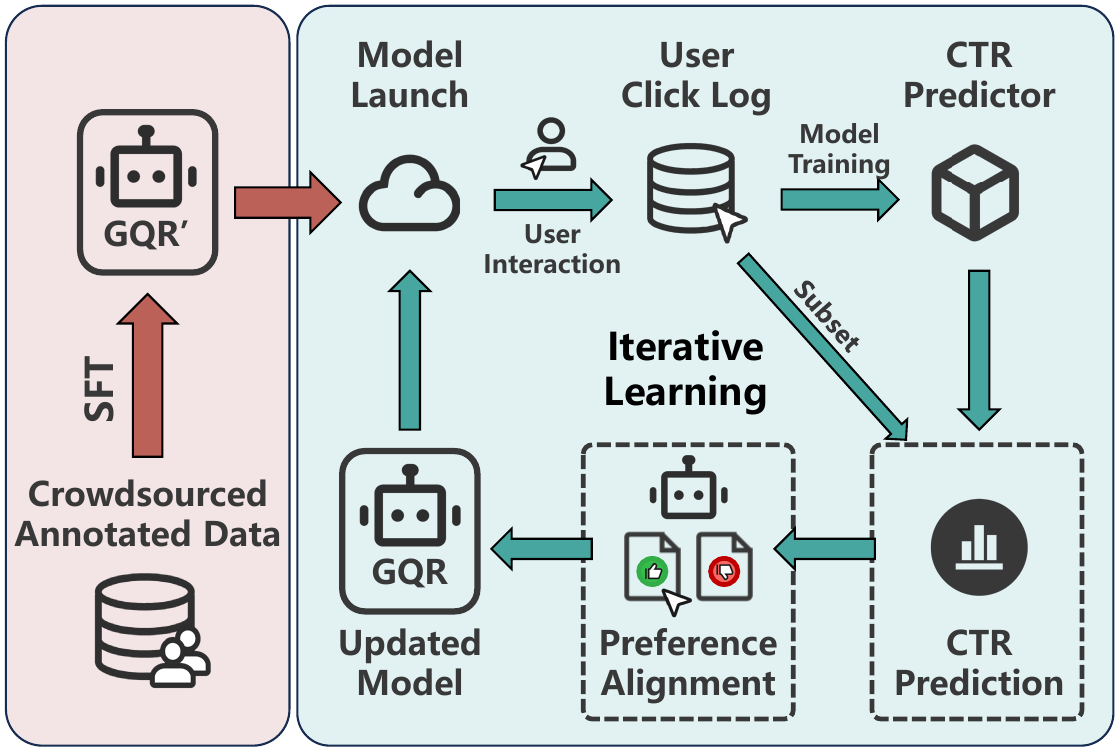} %
    \caption{The overall framework of enterprise GQR system.}
    \label{fig:pipeline}
    \vspace{-1em}
\end{figure}

\begin{table*}[]
\caption{Evaluation of CTR prediction models. Boldface denotes the highest score. $\uparrow$ denotes that larger score indicates better performance, while $\downarrow$ means the opposite. Note that we do not present (+pid) and (+ctx) results for Task 3 because it only requires to generate one query at one time. Details of the compared models are elaborated in Appendix \ref{sec:baselines}.}
\resizebox{0.8\linewidth}{!}{
\begin{tabular}{l|ccc|ccc|ccc}
\hline
\multirow{2}{*}{} & \multicolumn{3}{c|}{Task1}                            & \multicolumn{3}{c|}{Task2}                            & \multicolumn{3}{c}{Task3}                               \\ \cline{2-10} 
                  & AUC$\uparrow$             & Logloss$\downarrow$         & $\text{diff}_{ctr}\downarrow$      & AUC$\uparrow$             & Logloss$\downarrow$         & $\text{diff}_{ctr}\downarrow$      & AUC$\uparrow$              & Logloss$\downarrow$          & $\text{diff}_{ctr}\downarrow$      \\ \hline
BERT              & 0.7499          & 0.1785          & 5.134e-4          & 0.7678          & 0.1095          & 8.432e-4          & \textbf{0.7141}  & 0.1918  & 3.293e-4          \\ \hline
BERT+pid          & 0.7561          & 0.1771          & 5.132e-4          & 0.7736          & 0.1084          & 8.411e-4          & \textbackslash{} & \textbackslash{} & \textbackslash{}  \\ \hline
BERT+ctx          & 0.7559          & 0.1772          & 5.139e-4          & 0.7766          & 0.1082          & 8.417e-4          & \textbackslash{} & \textbackslash{} & \textbackslash{}  \\ \hline
RoBERTa           & 0.7493          & 0.1785          & 5.178e-4          & 0.7611          & 0.1099          & 8.433e-4          & 0.7121           & 0.1925           & 3.322e-4          \\ \hline
RoBERTa+pid       & 0.7528          & 0.1778          & 5.165e-4          & 0.7678          & 0.1093          & 8.429e-4          & \textbackslash{} & \textbackslash{} & \textbackslash{}  \\ \hline
RoBERTa+ctx       & 0.7540          & 0.1773          & 5.166e-4          & 0.7701          & 0.1087          & 8.401e-4          & \textbackslash{} & \textbackslash{} & \textbackslash{}  \\ \hline
ALBERT            & 0.7518          & 0.1781          & 5.119e-4          & 0.7724          & 0.1084          & 8.368e-4          & 0.7140           & \textbf{0.1918}           & \textbf{3.289e-4} \\ \hline
ALBERT+pid        & \textbf{0.7565} & \textbf{0.1766} & 5.114e-4          & 0.7761          & 0.1086          & 8.369e-4          & \textbackslash{} & \textbackslash{} & \textbackslash{}  \\ \hline
ALBERT+ctx        & 0.7560          & 0.1772          & \textbf{5.113e-4} & \textbf{0.7799} & \textbf{0.1080} & \textbf{8.357e-4} & \textbackslash{} & \textbackslash{} & \textbackslash{}  \\ \hline
\end{tabular}
}
\label{eva_ctr}
\end{table*}

\subsection{User Initiative Alignment}
\label{sec:cor}
Traditional Query Recommendation (TQR) systems relying on users' historical interactions require frequent updates with new user-initiated queries to keep pace with evolving user search patterns. In contrast, a GQR system does not draw from a constantly updated query database to retrieve queries or learning models, instead, all recommended queries are generated by LLMs token-by-token.
Despite the ease of use and flexibility of this end-to-end generation manner, the complete decoupling from user-initiated queries makes the LLM thoroughly unaware of the user’s proactive search preferences, potentially incurring suboptimal performance in some GQR tasks.
Therefore, in this section, we introduce how to align LLMs with users' proactive search preferences.
Essentially, we aim to introduce a Co-Occurrence Retrieval (COR) channel to retrieve co-occurrence input queries for each user query as the reference for LLM augmentation, as depicted in Figure~\ref{fig:psp}. We follow the steps below:

\textbf{(1)} We first collect large-scale search query logs from our system. 
Next, we extract a set of query triples $(q, q_{next}, c)$, where $q$ represents a query, $q_{next}$ is the user-initiated query that follows $q$ within the same search session, and $c$ denotes the frequency of this occurrence, aggregated over the entire query log data for all users. Based on these query triples, we construct a co-occurrence query dictionary $\mathcal{D}_{coo}$, where each key is a search query $q$, and the corresponding value is a list of immediate successive queries $\mathcal{Q}_q={q_{next}^1,\ldots,q_{next}^{N_q}}$, sorted in descending order by frequency.

\textbf{(2)} Since search queries typically follow a long-tail distribution, co-occurrence queries can be extremely sparse. To predict unseen queries, we develop a semantic matching model based on ERNIE \cite{zhang2019ernie} architecture and SimCSE \cite{gao2021simcse} framework. After training, the learned embeddings for all queries are deployed using an Approximate Nearest Neighbor (ANN) system.
During the inference process, once $\lvert \mathcal{Q}_q\rvert < n$ ($n$ is a pre-defined threshold value) given a search query $q$, $N_s$ nearest queries are retrieved from the ANN system and then used to retrieve co-occurrence queries from $\mathcal{D}_{coo}$, followed by a sort algorithm to obtain the final candidates $\mathcal{Q}_c^q$.

\textbf{(3)} Despite having co-occurrence queries $\mathcal{Q}_c^q$ for each search query $q$, they tend to be noisy and may not always offer valuable information about proactive search preferences.
To empower the LLM to selectively extract useful information from $\mathcal{Q}_c^q$, we fill $\mathcal{Q}_c^q$ into the prompt as additional side information (as stated in Template \ref{box:prompt}).
During the data annotation process for the initial SFT step, we instruct annotators (or prompt GPT) to uniformly annotate $0\sim N_{ref} (N_{ref}\leqslant N)$ recommended queries that are pertinent to $\mathcal{Q}_c^q$, with the rest being irrelevant. 
These lenient annotation criteria enable the model to leverage knowledge from $\mathcal{Q}_c^q$ to varying degrees, while the balance between the co-occurrence information and LLM inherent knowledge is regulated through CTR-alignment, as discussed in Section \ref{sec:ctr_align}.



\subsection{System Deployment}
\subsubsection{Periodic Update}
Since users' search preferences evolve over time, it is essential to periodically update the entire system when deploying an enterprise-level GQR system. Specifically, the co-occurrence system, CTR prediction model, and CTR alignment on the LLM need to be regularly updated to effectively cater to users' changing click preferences and improve the overall user experience. The overall learning framework or GQR is depicted in Figure~\ref{fig:pipeline}.

\subsubsection{Exploratory Strategies}
GQR recommended queries in a generation paradigm instead of retrieval and ranking. As a result, exploratory strategies are especially crucial for GQR to obtain comprehensive feedback from users, thereby continuously improving the system's performance. 
More specifically, we adopt three strategies to improve the exploration of our system. Due to space limitations, the details are placed in Appendix~\ref{sec:exploratory}.

\section{Experiments}
In this section, we conduct extensive experiments to evaluate the proposed GQR framework and answer the following questions:
\begin{itemize}
    \item \textbf{RQ1:} Can the proposed CTR predictor serve as a reliable estimator of real CTR?
    \item \textbf{RQ2:} Does our proposed GQR approach improve the estimated CTR of generated queries?
    \item \textbf{RQ3:} What is the level of quality exhibited by the generated queries?
    \item \textbf{RQ4:} How does each module contribute to the model performance?
    \item \textbf{RQ5:} Does the proposed framework perform well in online service?
\end{itemize}

\subsection{Experimental Setup}


\subsubsection{Tasks and Datasets}
We evaluate our methodology on three query recommendation tasks in Baidu Conversational AI assistant: 
\begin{itemize}
    \item \textbf{(T1) Query suggestion for general user needs:} This task is to provide multiple recommended queries (three queries in our setting) to users for preference elicitation after answering each user query \cite{shen2024enhancing}, which is a technology widely adopted in many AI assistant products, such as Gemini\footnote{https://gemini.google.com/}, Perplexity\footnote{https://www.perplexity.ai/}, and Yiyan\footnote{https://yiyan.baidu.com/},  etc. 
    \item \textbf{(T2) Query snippet generation for creative writing:}
    This task follows a choice-based query clarification format \cite{ZamaniGeneratingClarifyingQuestions}, where users are presented with a guiding sentence and multiple ($\sim 8$) clickable query snippets for each creative writing query (e.g., ``shorter'', ``more creative'', ``more elegant''), allowing them to refine their writing requirements.
    \item \textbf{(T3) Query hint in the search box:}
    This task aims to suggest a query in the search box for users based on the interaction history of the current session, which is also adopted by many commercial search systems.
\end{itemize}
Snapshots of the three components are illustrated in Figure~\ref{fig:showcases}. For each LLM calling in the three tasks, the current query, the AI response to the current query, and the historical queries within the current session are used to construct prompts. 
In task 2, apart from RQs, a guiding sentence, a prefix and a suffix (for forming a complete query) are generated as auxiliary texts.
We use real large-scale search logs from the Baidu Search AI assistant for model training and evaluation.
Specifically, for each task, we first construct a dataset for CTR predictor training via sampling 14 days of click logs from the initial SFT baseline. Afterwards, we randomly sample 10,000 queries from the 15th-21st days of search logs as the training set for CTR alignment, with 10,000 queries from the 22nd day as the test set.

\subsubsection{Evaluation Protocols and Baselines}
Due to space limitations, we place this part in Appendix~\ref{sec:metric} and \ref{sec:baselines}.

\subsection{Evaluation of CTR Predictors (RQ1)}
The performance of CTR predictors is the footstone of our GQR system, therefore, in this section, we compare the results of various CTR predictors.
From Table~\ref{table:ctr}, we have the following observations: (1) In general, models with position identification perform better than those without it, highlighting the importance of debiasing position information when predicting the CTR of a query.
(2) In Task 2, (+ctx) normally increase the scores, while in Task 1, the improvement is not significant, and even brings slight decay in certain metrics. We guess that it is because in Task 2 (short facet recommendation), previous exposed facets have more impacts on the click probability of current query. However, in Task 1, CTR of queries is relatively independent, and introducing context queries as input leads to more irrelevant noise and thus impairs the performance. (3) The differences in predicted CTR scores and actual CTR in logs are quite small in scale (1e-4), demonstrating that AEC could be an effective indicator to predict the overall CTR of a given dataset.
Additionally, since ALBERT tends to have better performance compared with BERT and RoBERTa, we use ALBERT for subsequent experiments and evaluations.



\begin{table}[]
\caption{The overall performance of various models for GQR. Results of three categories of methods are presented: 1) Baselines without click alignment, 2) Preference alignment with click/non-click logs as chosen/rejected instances, 3) Preference alignment with predicted CTR scores. $\star$ denotes the compared baseline of other methods and  ``Impr.'' denotes the relative improvement. Details of the compared models and AEC score are elaborated in Appendix \ref{sec:settings}.}
\resizebox{1.0\linewidth}{!}{
\begin{tabular}{l|l|cc|cc|cc}
\hline
\multirow{2}{*}{Cate.} & \multicolumn{1}{c|}{\multirow{2}{*}{Method}} & \multicolumn{2}{c|}{Task1} & \multicolumn{2}{c|}{Task2} & \multicolumn{2}{c}{Task3} \\ \cline{3-8} 
                       & \multicolumn{1}{c|}{}                        & AEC          & Impr.       & AEC          & Impr.       & AEC          & Impr.      \\ \hline
\multirow{2}{*}{Baselines}  & few-shot                                     & 0.1241       &0.65\%& 0.1833       &-19.75\%& 0.0679       &-0.43\%\\ \cline{2-8} 

                       & $\text{SFT}^\star$                                          & 0.1233       &0.00\%& 0.2284       &0.00\%& 0.0682       &0.00\%\\ \hline

\multirow{4}{*}{CLK}   & +$\text{SFT}_{clk}$                          & 0.1310       &6.24\%& 0.2290       &0.26\%& 0.0703       &3.08\%\\ \cline{2-8} 

                       & +$\text{KTO}_{clk}$                          & 0.1210       &-1.87\%& 0.2250       &-1.49\%& 0.0705       &3.37\%\\ \cline{2-8} 

                       & +$\text{SimPO}_{clk}$                        & 0.1217       &-1.30\%& 0.2311       &1.18\%& 0.0714       &4.69\%\\ \cline{2-8} 

                       & +$\text{DPO}_{clk}$                          & 0.1227       &-0.49\%& 0.2303       &0.83\%& 0.0731       &7.18\%\\ \hline

\multirow{5}{*}{CTR}   & +$\text{SFT}_{ctr}$                          & 0.1681       &36.33\%& 0.2510       &9.89\%& 0.0751       &10.12\%\\ \cline{2-8} 

                       & +$\text{KTO}_{ctr}$                          & 0.1850       &50.04\%& 0.2518       &10.25\%& 0.0826       &21.11\%\\ \cline{2-8} 

                       & +$\text{SimPO}_{ctr}$                        & 0.1952       &58.31\%& 0.2683       &17.47\%& 0.0915       &34.16\%\\ \cline{2-8} 

                       & +$\text{DPO}_{clk}$                          & 0.2510       &103.57\%& 0.3208       &40.46\%& 0.1122       &64.52\%\\ \cline{2-8} 

                       & +Iter $\text{DPO}_{clk}$                     & 0.2711       &119.87\%& 0.3566       &56.13\%& 0.1213       &77.86\%\\ \hline
\end{tabular}}
\label{table:ctr}
\end{table}

\subsection{Evaluation of Preference Alignment (RQ2\&RQ3)}
\subsubsection{Evaluation of CTR Improvements}

We use AEC (as mentioned in Section \ref{sec:metric}) to evaluate the performance of various alignment techniques for GQR in Table~\ref{eva_ctr} and Table~\ref{table:ctr}. We find the following observation:

\begin{itemize}
    \item The performance of SFT with human-annotated training data and in-context learning using few-shot examples is similar in Task 1 and Task 3, suggesting that utilizing more annotated data for SFT does not necessarily enhance recommendation performance. This implies that the strength of instruction-following capabilities is not a decisive factor for generative recommendation tasks. This is reasonable because even with high-quality annotations for GQR tasks, there remains a significant discrepancy between the annotators' preferences and the users' actual click preferences.
    \item Overall, aligning preference directly with click samples can moderately improve the AEC of the generated recommendation queries, although the effectiveness is not always guaranteed. For instance, using click samples for SFT can slightly improve AEC from 0.1233 to 0.1310 in Task 1, representing a $6.24\%$ increase. However, constructing positive and negative samples for preference alignment based solely on clicked/non-clicked samples might degrade performance, as seen in the case of KTO\(_{clk}\) (-1.87\%), SimPO\(_{clk}\)(-1.30\%), DPO\(_{clk}\)(-0.49\%) for Task 1, and KTO\(_{clk}\) (-1.49\%) for Task 2. This is due to the noisy and random nature of user click and non-click signals, which adversely affects preference learning. Click-based alignment consistently improves performance in Task 3 compared to the other two tasks. This could be attributed to Task 3 recommending only one query, resulting in less noise in the chosen and rejected responses compared to the other tasks.
    \item Methods that perform preference alignment based on predicted CTR signals consistently yield improvements. Specifically, employing high-CTR filtered positive samples for SFT results in noticeable enhancements, such as an increase in AEC from 0.1233 to 0.1681 in Task 1, marking a $36.33\%$ improvement. This indicates that filtering based on CTR scores predicted by CTR models is more accurate than direct filtering based on click signals, thereby reducing noise and variance. When low-CTR rejected responses are added as negative samples, methods like KTO\(_{ctr}\), SimPO\(_{ctr}\), and DPO\(_{ctr}\) exhibit significant improvements. For example, DPO\(_{ctr}\) in Task 1 shows a dramatic CTR increase from 0.1233 to 0.2510, a $103.57\%$ improvement. This demonstrates that preference alignment is more effective than SFT alone, and understanding the patterns of negative samples is crucial for the model. Among these, pair-sample-based methods such as SimPO\(_{ctr}\) and DPO\(_{ctr}\) outperform KTO\(_{ctr}\), suggesting that pair samples facilitate a better understanding of user preferences. Furthermore, Iterative DPO\(_{clk}\) outperforms DPO\(_{clk}\), achieving the best AEC of 0.2711 (Task 1), 0.3566 (Task 2), 0.1213 (Task 3), indicating that multiple iterations can further enhance the performance.
\end{itemize}

\subsubsection{Evaluation of Query Qualities}
\begin{figure}[!htb]
    \centering
    \begin{minipage}[b]{.49\linewidth}
        \centering
        \includegraphics[width=\linewidth]{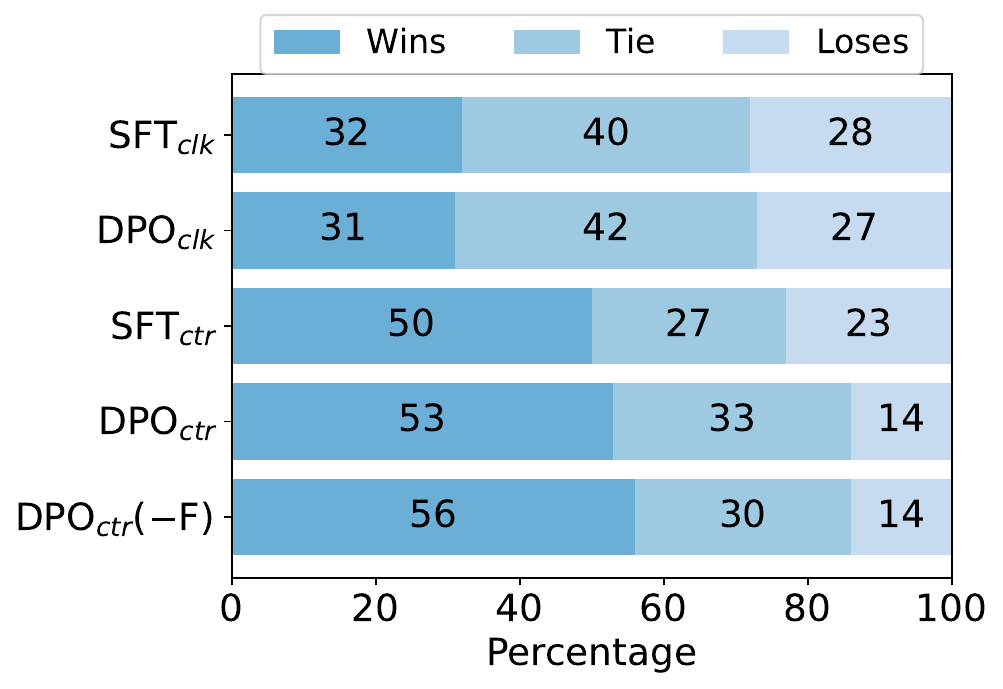}
        \subcaption{Task 1 Relatedness}
        \label{fig:task1rel}
    \end{minipage}
    \hfill
    \begin{minipage}[b]{.49\linewidth}
        \centering
        \includegraphics[width=\linewidth]{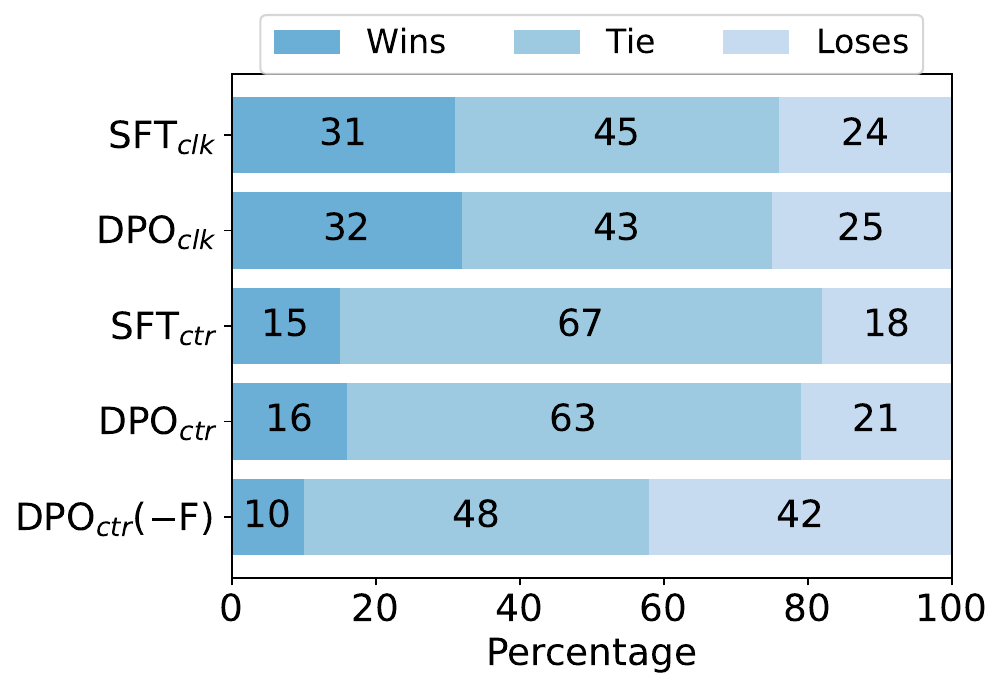}
        \subcaption{Task 1 Semantic Exclusivity}
        \label{fig:task1exc}
    \end{minipage}
    
    \begin{minipage}[b]{.49\linewidth}
        \centering
        \includegraphics[width=\linewidth]{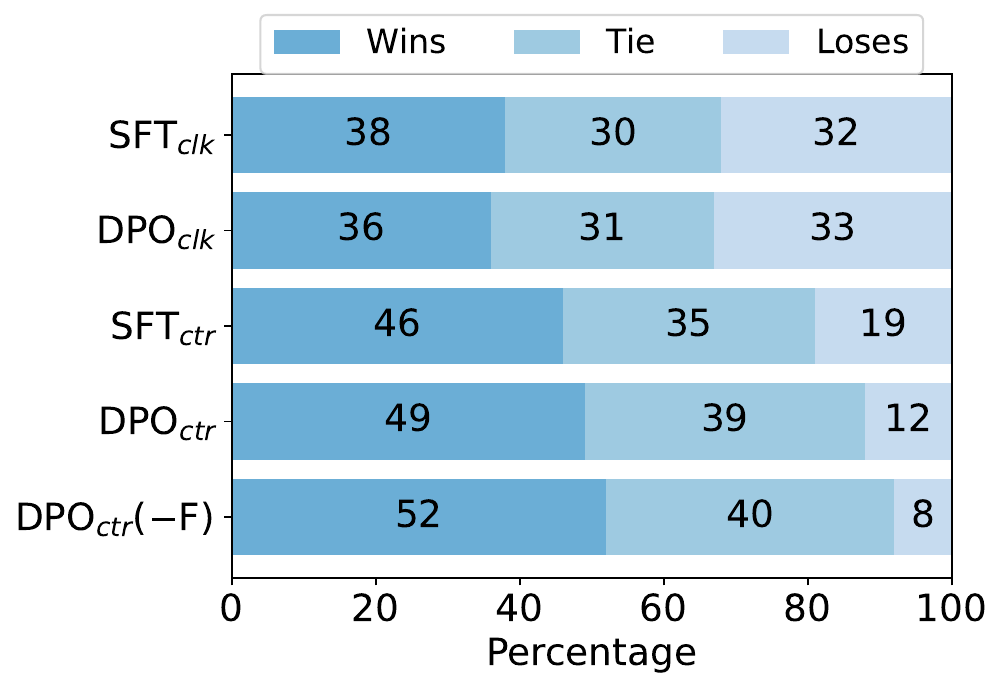}
        \subcaption{Task 2 Relatedness}
        \label{fig:task2rel}
    \end{minipage}
    \hfill
    \begin{minipage}[b]{.49\linewidth}
        \centering
        \includegraphics[width=\linewidth]{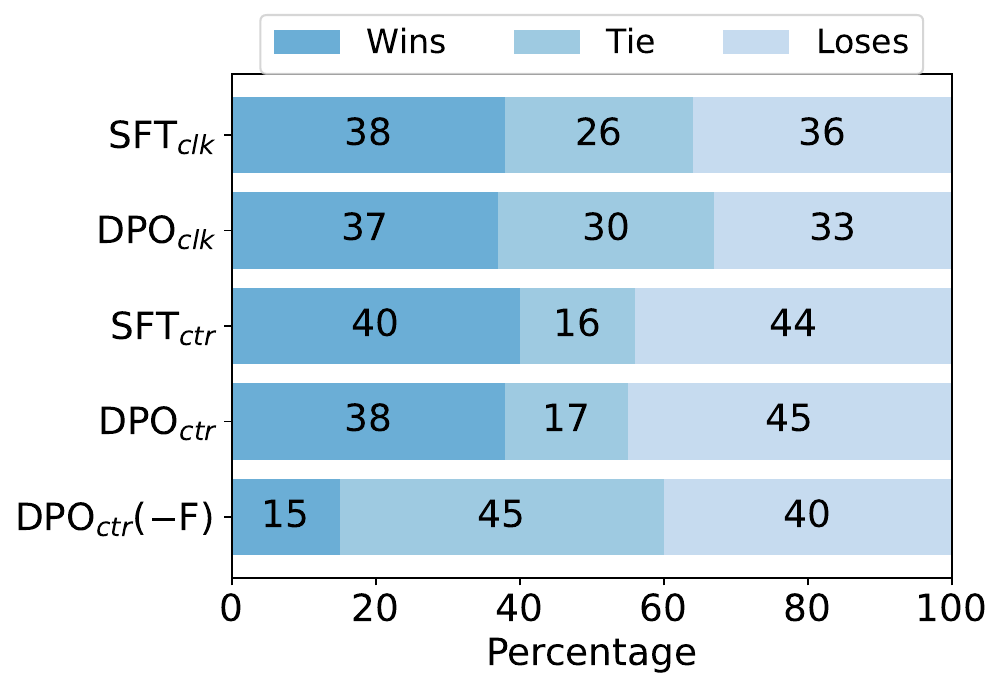}
        \subcaption{Task 2 Semantic Exclusivity}
        \label{fig:task2exc}
    \end{minipage}
\caption{Evaluation of Query Qualities.}
\label{fig:quality}
\end{figure}

In this section, we evaluate the quality of generated queries of different models in terms of Relatedness and Semantic Exclusivity (referring to Appendix \ref{sec:metric}). 
We set the initial SFT model without alignment as the baseline and give the Win-Tie-Loss (WTL) results of five methods: $\text{SFT}_{clk}$, $\text{DPO}_{clk}$, $\text{SFT}_{ctr}$, $\text{DPO}_{ctr}$, $\text{DPO}_{ctr}(-\text{F})$, for both Task 1 and Task 2. $(-\text{F})$ denotes removing the filter operation of overlapping queries as described in Section \ref{sec:alignment}.
From Figure~\ref{fig:quality}, we can observe that the results across both tasks follow a similar pattern. For relatedness, $\text{SFT}_{clk}$ and $\text{DPO}_{clk}$ show a slight improvement, whereas $\text{SFT}_{ctr}$, $\text{DPO}_{ctr}$, and $\text{DPO}_{ctr}(-\text{F})$ significantly enhance the relatedness. This trend aligns with the estimated CTRs, indicating a strong correlation between user click intent and the Relatedness between RQs and user queries.
Regarding semantic exclusivity, the WTL scores of $\text{SFT}_{ctr}$, $\text{DPO}_{ctr}$, and $\text{DPO}_{ctr}(-\text{F})$ remarkably decrease. This is reasonable since exclusivity and relatedness are inherently somewhat contradictory. Notably, the $\text{DPO}_{ctr}(-\text{F})$ method, which lacks sample filtering, has a substantially lower exclusivity score than the baseline. This could potentially affect user experience and lead to model degradation, highlighting the necessity of filtering operations.

\vspace{-1em}



\begin{figure}[h]
\centering
\includegraphics[width=0.49\linewidth]{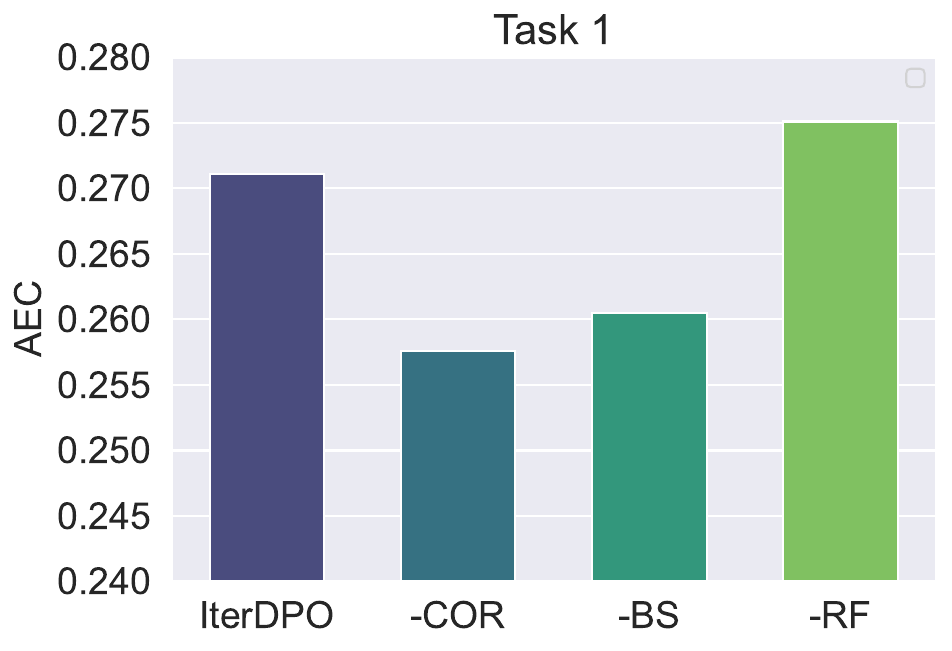} 
\includegraphics[width=0.49\linewidth]{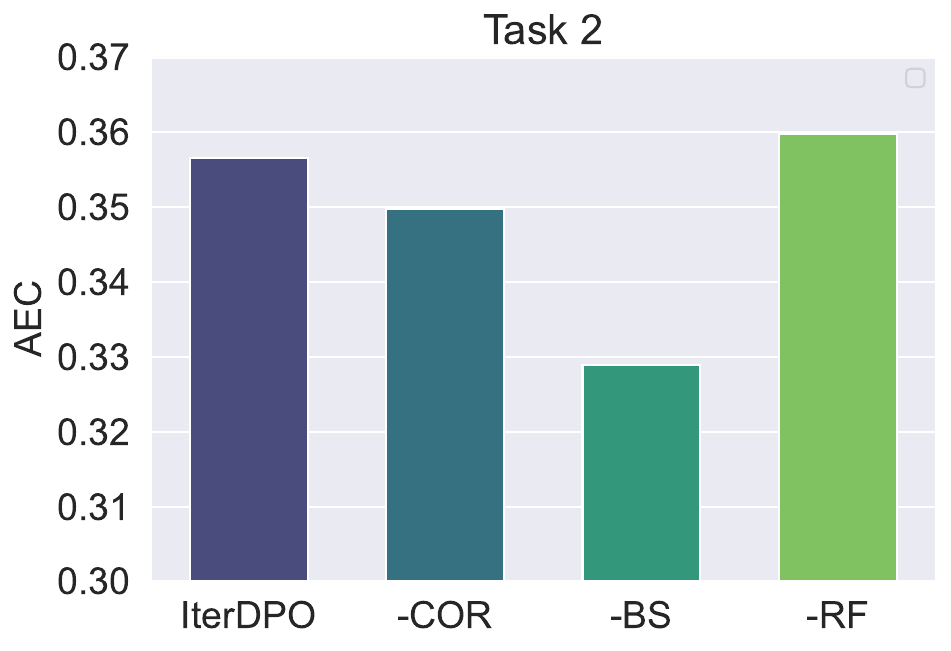} 
\vspace{-10pt}
\caption{Ablation study for key components}
\label{fig:abs}
\end{figure}

\subsection{Ablation Study (RQ4)}
In this section, we conduct the ablation analysis on Task 1 and Task 2 to confirm the effectiveness of each key component of the proposed approach. As depicted in Figure~\ref{fig:abs}, the notation \textbf{-COR} signifies removing the COR module (User Initiative Alignment) as elaborated in Section \ref{sec:cor}, \textbf{-BS} indicates the substitution of beam search sampling for chosen responses with random sampling, and \textbf{-RF} represents the absence of filtering semantically similar queries. It can be observed that \textbf{-COR} notably reduces performance,
highlighting the significance of aligning LLM with proactive search preference. Similarly, \textbf{-BS} also undermines the performance, particularly in Task 2. The reason is that searching high-CTR queries sequentially is more likely to yield responses with higher CTR. Since in Task 2, the LLM is prompted to generate more query facets ($\sim$ 8), employing beam search has a more pronounced effect. Conversely, \textbf{-F} marginally improves the AEC, but introduces non-negligible adverse effects on response quality, as depicted in Figure~\ref{fig:quality}.

\begin{figure}[h]
\centering
\includegraphics[width=0.8\linewidth]{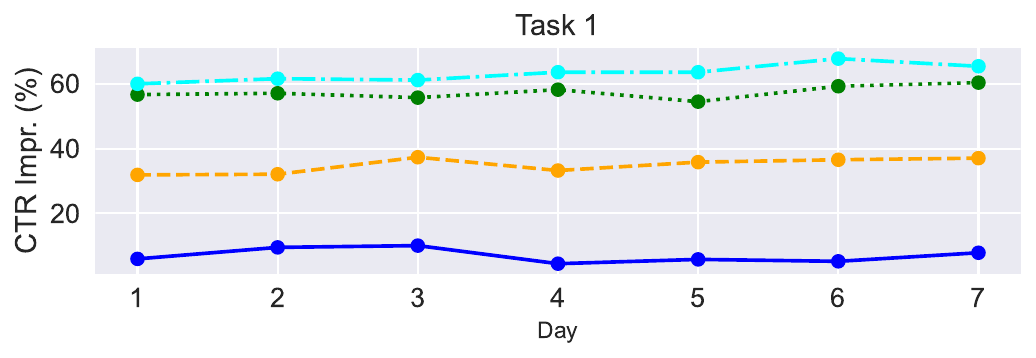} 
\hfill
\includegraphics[width=0.8\linewidth]{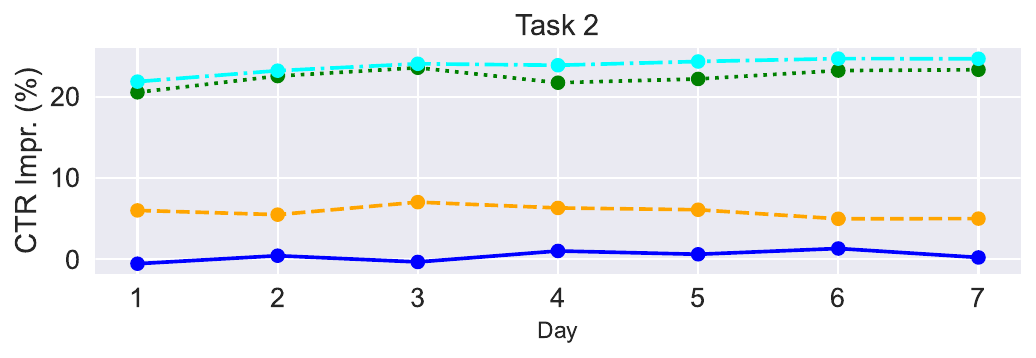} 
\hfill
\includegraphics[width=0.8\linewidth]{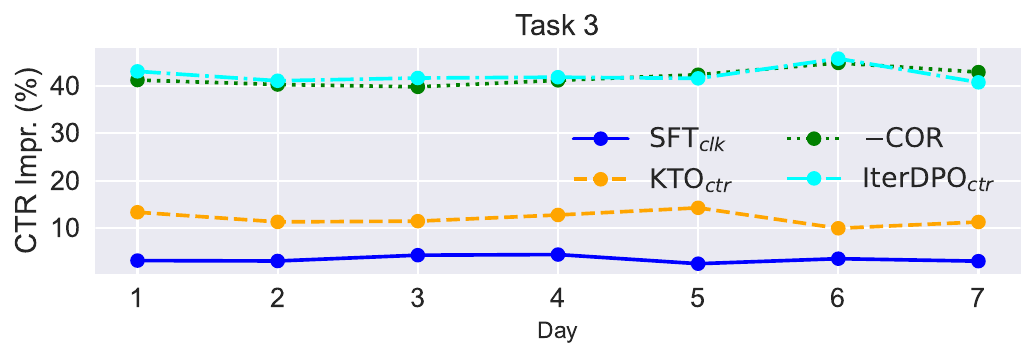} 
\caption{Results from Online A/B test during 7 consecutive days.}
\label{fig:online_res}
\end{figure}

\subsection{Online Evaluation (RQ5)}
We conducted a consecutive 7-day A/B test on the Baidu Search AI Assistant service. Figure~\ref{fig:online_res} shows the online results of several methods. In consideration of the company’s policy, we did not show the actual CTR values, but rather the percentage increase of CTR for each method compared to the baseline (the baseline is the initial SFT model). From the chart, we have the following observations. 
\begin{itemize}
    \item On the whole, the trend is relatively consistent with the estimated CTR, implying that AEC can serve as an offline indicator for online performance, verifying the efficacy of our GQR framework, which achieves $63\%,24\%,42\%$ improvements on average in the three tasks, respectively.
    \item In the case where the actual CTR of the model ($\text{SFT}_{clk}$) is relatively close to the baseline, the percentage increase in the actual CTR and the estimated CTR (AEC) are quite similar. Conversely, when there is a large gap between the actual CTR and the baseline, the difference between the percentage increase of the actual CTR and AEC is significant. This is reasonable because the CTR model trained on user click logs based on the baseline model has difficulty accurately estimating outliers with extremely high CTR. 
    \item Over time, the CTR improvement ratios of $\text{Iter\_DPO}_{ctr}$ show an upward trend, suggesting that significantly improving the CTR also enhances the user experience.
\end{itemize}



\section{Conclusion}
In this study, we introduce the Generative Query Recommendation (GQR) paradigm to address diverse query recommendation tasks in search engines, which aligns LLMs with user click preference.
Our method comprises a general prompt framework for LLM-based query recommendation tasks, a CTR-alignment framework to enhance user engagement, and alignment with proactive search preference using co-occurrence retrieval augmentation.
Implementing GQR in Baidu’s conversational search services led to notable enhancements in CTR, 
moving beyond traditional models that rely heavily on historical search logs.
\bibliographystyle{ACM-Reference-Format}
\bibliography{main}

\appendix
\section{Details of DPO}
\label{sec:DPO}
Direct Preference Optimization (DPO) is an approach for fine-tuning language models to align with human preferences without requiring reinforcement learning. Unlike Reinforcement Learning from Human Feedback (RLHF), which involves explicitly training a reward model and using reinforcement learning techniques like Proximal Policy Optimization (PPO), DPO directly optimizes the model to increase the probability of preferred responses over dispreferred ones.

The process of DPO starts with supervised fine-tuning (SFT) of a pre-trained LM to obtain the initial SFT model $\pi^{SFT}$ that is suitable for our query generation task. Then, instead of explicitly training a reward model and performing reinforcement learning as in RLHF, DPO directly optimizes the model using preference data, as described in Section \ref{sec:alignment}. Given a dataset of preference pairs $\mathcal{D}=\{\left(x_{i},y_{i}^{w},y_{i}^{l}\right)\}_{i=1}^{N}$, where ${y}_{i}^{w}$ represents the chosen response and $y_{i}^{l}$ represents the rejected response for input $x_{i}$, DPO fine-tunes $\pi^{SFT}$ by maximizing the likelihood of preferred responses relative to less preferred ones.

To achieve this, DPO introduces an implicit reward function $r_{\theta}(x,y)$, assuming that the probability ratio of the preferred and dispreferred responses follows:
\begin{equation}
\frac{\pi_{\theta}(y^{w}|x)}{\pi_{\theta}(y^{l}|x)}=e^{r_{\theta}(x,y^{w})-r_{\theta}(x,y^{l})}
\end{equation}
where $\pi_{\theta}$ is the fine-tuned model we aim to optimize. Unlike RLHF, where a separate reward model is trained and used within a RL framework, DPO directly optimizes $\pi_{\theta}$ by leveraging the reference model $\pi_{ref}$
  (typically the SFT model $\pi^{SFT}$) as a baseline.
The optimization objective in DPO is derived from the likelihood ratio between the fine-tuned model and the reference model:
\begingroup
\tiny
\begin{equation}
\mathcal{L}_{\mathrm{DPO}}(\pi_{\theta};\pi_{\mathrm{ref}})=-\mathbb{E}_{(x,y_{w},y_{l})\sim\mathcal{D}}\left[\log\sigma\left(\beta\log\frac{\pi_{\theta}(y_{w}\mid x)}{\pi_{\mathrm{ref}}(y_{w}\mid x)}-\beta\log\frac{\pi_{\theta}(y_{l}\mid x)}{\pi_{\mathrm{ref}}(y_{l}\mid x)}\right)\right]
\end{equation}
\endgroup
where $\sigma(\cdot)$ is the sigmoid function, and $\beta$ is a temperature parameter controlling the sharpness of preference modeling. This objective encourages 
$\pi_{\theta}$ to increase the probability of generating preferred responses while staying close to the reference model $\pi_{ref}$ to ensure stability.
By avoiding explicit reward modeling and reinforcement learning, DPO offers a more stable and efficient approach to aligning language models with human preferences.

\section{Experimental Settings}
\label{sec:settings}
\subsection{Evaluation Protocols}
\label{sec:metric}
To assess the performance of CTR prediction, we exploit the most commonly-used AUC (Area Under the ROC) \cite{hanley1982meaning} and logloss to evaluate the model. Furthermore, since the CTR predictor is utilized to estimate the CTR of responses generated by LLMs, we also propose a new metric to evaluate its capability to fit actual online CTR:
\begin{equation}
    \text{diff}(CTR_\theta, CTR_\text{real}) = \frac{1}{K}\sum_i^K\lvert\frac{p^\theta_i \overline{p}^\text{real}_i}{\overline{p}^\theta_i}- p^\text{real}_i)\rvert,
\end{equation}
where $p_i^\theta$ is the Average Estimated CTR (AEC) of each instance in the testing set $i$, and $p^\text{real}_i$ is the click ratio on the testing set $i$, $K$ is the number of testing sets, $\overline{p}^\text{real}_i$ and $\overline{p}^\theta_i$ are average scores of $p_i^\theta$ and $p^\text{real}_i$, respectively. The $K$ testing sets are randomly sampled from logs of $K$ days, respectively.

Common metrics to evaluate the effectiveness of query generation are term overlap (Precision, Recall, and F1), exact match, or similarities between predicted texts and human-annotated ground truth. However, in our setting, there is no ground truth response for each query and consequently, we evaluate the GQR performance from two dimensions: CTR and query quality. In terms of CTR, we adopt 1) AEC on responses of the testing set using our CTR predictor for offline evaluation, and 2) actual CTR in online A/B testing.
In terms of query quality, we conduct human evaluation according to two criteria in Baidu Search for fine-grained assessments:

\textbf{Relatedness}. It focuses on the Relatedness between RQs and the user query. It has three levels: 1) \textit{High}: The recommended query closely aligns with the user’s query in terms of topic and intent, and it supplements the user’s query with additional, relevant information; 2) \textit{Moderate}: The recommended query is somewhat related to the user’s query but may not directly fulfill the user’s original intent, offering content that is more of a tangential relevance; 3) \textit{Low}: The recommended query has low relevance to the user’s query, with significant differences in topic and intent.

\textbf{Semantic Exclusivity}. It focuses on three categories of exclusivity of RQs: 1) \textit{Exclusivity with the user query}: This aspect measures whether the RQ is semantically equivalent to the user query, ensuring that it offers a unique perspective or new direction.
2) \textit{Exclusivity with AI response}: This dimension evaluates how the recommended query influences the uniqueness of the AI generated content or answers, ensuring that it contributes novel information.
3) \textit{Diversity of RQs}: This aspect assesses the degree to which the recommended query is distinct from other recommended queries, ensuring that each query contributes uniquely to the overall set.

\subsection{Baselines}
\label{sec:baselines}
For CTR prediction, we evaluate three BERT-like architectures: BERT \cite{devlin2018bert}, RoBERTa \cite{liu2019robertarobustlyoptimizedbert}, and ALBERT \cite{lan2020albertlitebertselfsupervised}. 
We assess three alternatives for each architecture: 1) Concatenate $q_{user}$ with $q_k$, separated by a <SEP> token and segment embeddings; 2) additionally concatenate a position identification ``$\_k$'' with $q_k$ to allow the model to distinguish the position bias, denoted as (+pid); 3) Input structured as Figure~\ref{fig:ctr_model}, facilitating the model to capture context information and position bias, indicated as (+ctx).

For GQR models, we evaluate three categories of approaches: 
\begin{itemize}
    \item \textbf{Baselines}: We implement two baselines: 1) \textit{Few-shot}: LLM prompted with five selected examples. 2) \textit{SFT}:  fine-tune LLM on around 500 human-annotated instances to obtain $\pi^{SFT}_\phi$. 
    \item \textbf{LLM aligned with click signals}: This category of methods directly aligns LLMs with click signals. Specifically, we collect prompt-response pairs from exposure logs and extract those with click behaviors. $\text{SFT}_{clk}$ denotes supervised fine-tuning $\pi^{SFT}_\phi$ on $N_{pos}$ instance in which one of the generated queries is clicked. $\text{KTO}_{clk}$ is trained on the same $N_{pos}$ clicked instance, and $N_{neg}$ randomly sampled unclicked instance. $\text{SimPO}_{clk}$ and $\text{DPO}_{clk}$ require preference pairs, so we use the online response of each clicked instance as the chosen response, and then
    re-generate responses for each prompt, and select the response which does not include the clicked query in the chosen response as the rejected response. $N_{pos}=N_{neg}=10000$ for all models. 
    \item \textbf{LLM aligned with CTR signals}: $\text{KTO}_{ctr}$, $\text{SimPO}_{ctr}$, $\text{DPO}_{ctr}$ and $\text{IterDPO}_{ctr}$ follow the same steps in Algorithm \ref{iterDPO} to generate both chosen and rejected responses for each instance, whereas $\text{SFT}_{ctr}$ only use the chosen responses.
\end{itemize}
Note that all evaluated methods are incorporated with the COR module. Both Cate.2 and Cate.3 methods are trained based on the SFT baseline. For a fair comparison, all methods are trained based on the ERNIE-Speed\footnote{https://agicto.com/model/ERNIE-Speed-8K?status=2}, which is a high-performance foundation model released by Baidu Inc. Given that we found that the results marginally rely on the high-level capabilities (e.g., reasoning) of LLMs, we do not extensively compare the performance of different open-source LLMs, which can be put in our future work.

\begin{figure}[h]
\centering
\includegraphics[width=0.49\linewidth]{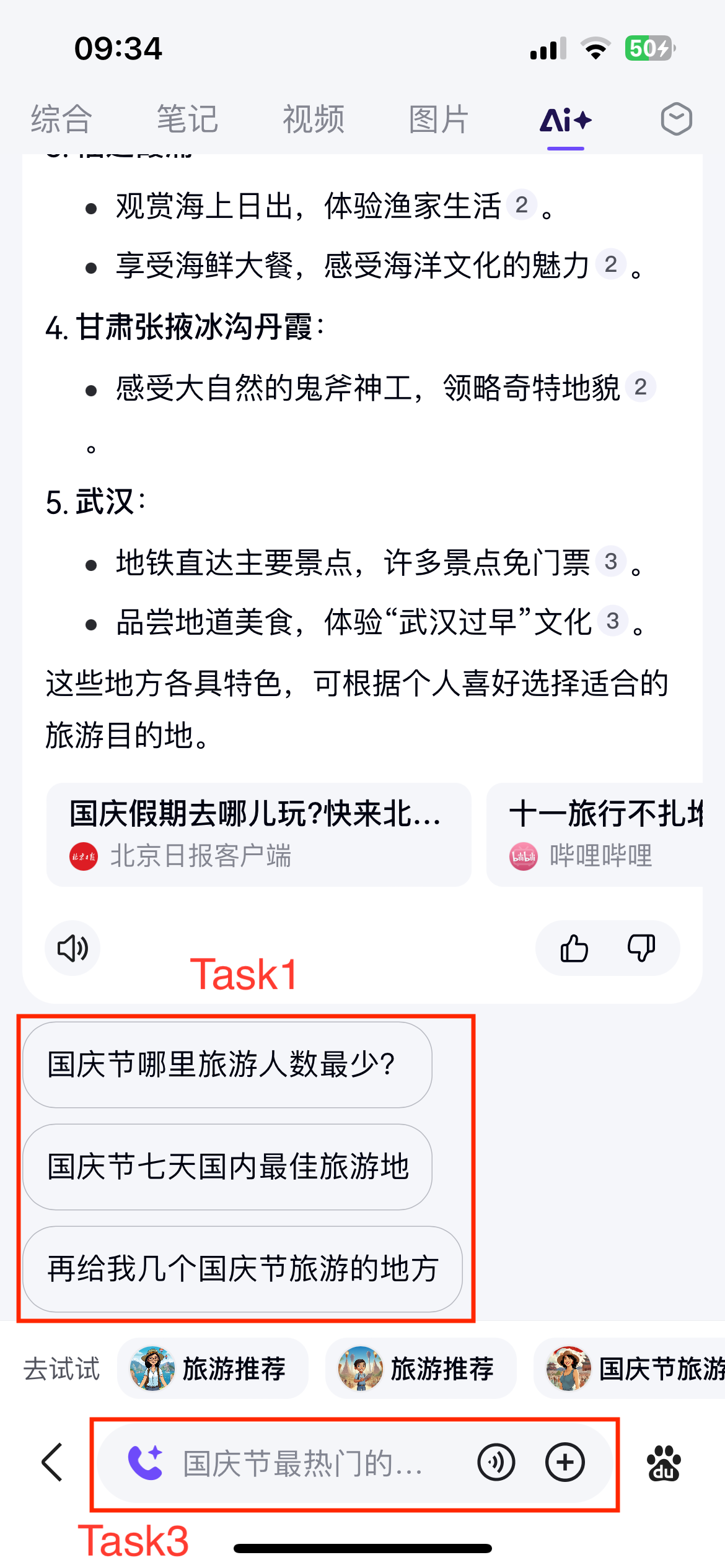} 
\includegraphics[width=0.49\linewidth]{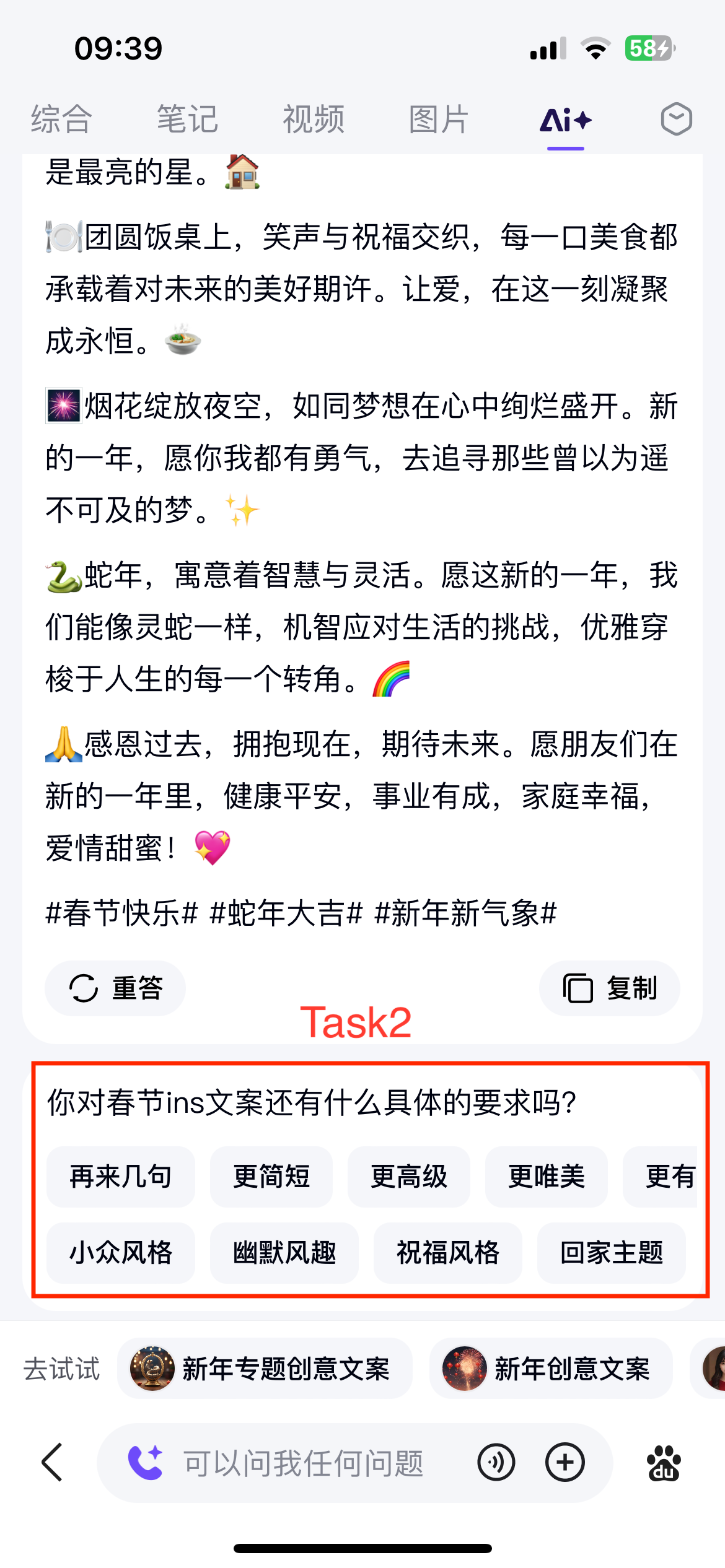} 
\caption{Illustration of the three query recommendation tasks in the Baidu Conversational Search System.}
\label{fig:showcases}
\end{figure}

\section{Exploratory Strategies}
\label{sec:exploratory}
This section elaborates the exploratory strategies adopted in our system:
\begin{itemize}
    \item \textbf{Diversity in Data Annotation:}
The formulation of RQs is significantly influenced by the intrinsic knowledge of LLMs and initial SFT data, potentially deviating from users' actual preferences. To mitigate the risk of getting stuck in local optimum, 
we augment the annotated SFT dataset $\mathcal{D}_{SFT}$ by prompting large LLMs (e.g., GPT) to rewrite the recommended queries with diverse styles and contents and obtain $\mathcal{D}_{SFT}^{aug}$. The two datasets are mixed to train the initial SFT model.
    \item \textbf{Diverse Beam Search for Generation of Training data:} In the process of candidate response generation (both chosen and reject responses), we incorporate diverse beam search \cite{vijayakumar2016diverse} techniques to ensure that a wide range of possible query variations are considered, increasing the possibility of generating distinctive responses.
    \item \textbf{Randomness in Online Decoding:}
When deploying the LLM online to provide query recommendations to users, we introduce randomness by tuning parameters such as temperature and diversity. This approach ensures that even for the same query, the model is capable of generating different answers over multiple interactions. By adopting this strategy, we aim to collect more comprehensive user click feedback, which is crucial for better capturing user preferences and subsequently improving the system’s recommendation accuracy. 
\end{itemize}

\section{A Brief Introduction of Baidu Conversational Search System}
Baidu Conversational Search System (BCSS) is complementary to the general Baidu search engine, which enables users to engage in multi-turn dialogues, allowing for more nuanced and context-aware interactions. It allows users to seek complex information that might require clarification or follow-up questions. BCSS employs a Retrieval-Augmented Generation (RAG) architecture, combined with various guiding and clarification components, to enhance the efficiency and accuracy of user interactions. We leverage the GQR framework proposed in this paper to implement the guiding and clarification functions. Three forms of query recommendation are presented in Figure~\ref{fig:showcases}, corresponding to the three tasks in our experiments.

\section{Related Work}
\label{sec:RelatedWork}

\subsection{Query Recommendation}
\label{sec:queryRec}
The primary objective of query recommendation (or query prediction) is to enhance the user search experience by aiding in the formulation of queries. It can be categorized into several tasks, such as query auto-completion \cite{2014QueryAutoCompletion,2015QueryAutoCompletion,SurveyofQueryAutoCompletion,ZamaniGeneratingClarifyingQuestions}, query suggestion \cite{uddin2018multitask,2019ClickFeedbackAwareQueryRecommendation,mustar2020usingBERTandBART,chen2020incorporatingbehavioralhypothesesquery,lee2024enhancedfacetgenerationllm,MiningExploratoryQueries_Dou} or query clarification \cite{ZamaniUserInteractions,2020clarifyingQuestion,2021FacetDrivenGenerationClarifyingQuestionsSearch,2022GeneratingClarifyingQuestionsWeb,ZamaniGeneratingClarifyingQuestions,ZamaniGeneratingClarifyingQuestions,pyatkin-etal-2023-clarifydelphi}, which have been extensively studied over the years.
Traditional methods primarily rely on heuristic rules and statistical models, leveraging query logs, term co-occurrence, and user behavior patterns to make query recommendation \cite{10.1145/1963405.1963424,10.1145/2505515.2505661,6816668,10.1145/3020165.3022129,10.1145/3077136.3080652}. These approaches, though effective, struggle with sparsity issues and adapting to evolving user intents.
Neural network-based methods address these limitations by leveraging distributed representations and sequence models \cite{10.1145/2806416.2806493,10.1145/3132847.3133010,10.1145/3397271.3401331,ZamaniConversationsClarifyingQuestions,2017NeuralLanguageModel4QAC,jaech2018personalizedlanguagemodelquery}. However, these methods require substantial training data and fail to handle long-tail, complicated or contextually-relevant settings.

With the rise of LLMs, query recommendation has increasingly shifted towards generative paradigms, addressing the aforementioned problem while enabling more flexible and context-aware suggestions \cite{2024MultimodalQuerySuggestion,ZeroShot-CQGen-4ConversationalSearch,2024KnowledgeAugmentedLLMPersonalizedQuerySuggestion,bacciu2024generatingqueryrecommendationsllms,2024SyntheticQueryGenerationUsingLLM4VirtualAssistants}. 
 For instance, \citet{lee2024enhancedfacetgenerationllm} propose integrating prior knowledge from smaller models with LLMs to generate refined query recommendations directly from user queries. \citet{MiningExploratoryQueries_Dou} adopts the ICL ability of the LLM to generate exploratory queries based on prompt engineering to enable a more diverse exploratory recommendation. Additionally, \citet{MultiTurnClarificationDou} leverage LLMs to cluster and select relevant query facets as recommendations, facilitating multi-turn interactions. 

Despite these advancements, existing studies primarily tackle these query-centric tasks independently \cite{10.1145/3209978.3210004,10.1145/2600428.2609614,10.1145/2600428.2609571,10.1145/1401890.1401995,10.1145/2433396.2433481,oard1998implicit,ZamaniUserInteractions,ZamaniGeneratingClarifyingQuestions}, limiting the generalizability of these approaches. Moreover, current LLM-based approaches fail to align model outputs with user feedback, leading to suboptimal online performance. In contrast, we propose a unified generative framework that integrates these query-centric recommendation tasks under a single paradigm, and devise a CTR alignment strategy along with an initiative alignment pipeline to improve user engagement.

\subsection{LLM-based Recommendation}
\label{sec:LLM_Rec}
The emergence of LLMs has demonstrated significant potential in the field of recommendations~\cite{Darec}. Broadly speaking, these approaches can be categorized into two primary types: discriminative LLMs and generative LLMs for recommendation \cite{wu2024survey}.

Discriminative LLMs in the recommendation domain primarily refer to models from the BERT series \cite{devlin2018bert,liu2019robertarobustlyoptimizedbert,lan2020albertlitebertselfsupervised}. Given their expertise in natural language understanding tasks, discriminative language models are often used as the backbone for various downstream tasks, including recommendation systems. Most existing work aligns the representations of pre-trained models like BERT with domain-specific data through fine-tuning \cite{qiu2021u,wu2021userbertcontrastiveusermodel,10.1145/3488560.3498495,10.1145/3485447.3511977}. Additionally, some research explores training strategies such as prompt tuning and adapter tuning \cite{10.1145/3383313.3412249,yang2021improvingconversationalrecommendationsystems,SHEN2023103139,10.1145/3539618.3591752}.
Unlike most discriminative model-based approaches that align the representation learned by LLMs to the recommendation domain, most generative model-based work translates recommendation tasks as natural language tasks, and then applies techniques such as in-context learning \cite{gao2023chatrecinteractiveexplainablellmsaugmented, He_2023LLMZeroShotRec,2023ChatGPTCapabilitiesRecSys,li2023gpt4recgenerativeframeworkpersonalized,sun2024snippetbasedconversationalrecommender}, prompt tuning \cite{2023TALLRec,10.1007/978-3-031-56063-7_42,yang2023palrpersonalizationawarellms,chu2023leveraginglargelanguagemodels}, and instruction tuning \cite{2022P5_Rec,cui2022m6recgenerativepretrainedlanguage,10.1145/3708882} to adapt LLMs to directly generate the recommendation results.
These approaches mainly focus on improving item-user correlations, while our GOR paradigm differs fundamentally by generating textual queries optimized for high user click preference.

\subsection{Preference Alignment for LLM}
\label{sec:PreferenceAlignment}
Many studies have focused on aligning LLMs with human preferences. Early efforts primarily revolved around Reinforcement Learning from Human Feedback (RLHF) \cite{ziegler2020finetuninglanguagemodelshuman, bai2022traininghelpfulharmlessassistant, NIPS2017_d5e2c0ad, NEURIPS2022_b1efde53}, where human value alignment is achieved by maximizing a scalar value derived from a reward model, combined with KL-regularization. This approach employs RL algorithms such as Proximal Policy Optimization (PPO) \cite{schulman2017PPO}. 
However, PPO often struggles with maintaining consistent response lengths and suffers from sudden reward drops. As a result, researchers have explored simpler, more efficient reward-free methods that directly use preference or ranking data to align LLMs with human preferences \cite{yuan2023rrhfrankresponsesalign,liu2024statisticalrejectionsamplingimproves,xu2024dposuperiorppollm}. 
Direct Preference Optimization (DPO) \cite{NEURIPS2023_DPO} has emerged as a popular alternative, as it can recover the same optimal policy as RLHF without the need to explicitly train a reward model. 
On the basis of DPO, IPO \cite{pmlr-v238-gheshlaghi-azar24a} introduces a method for learning from human preferences expressed in terms of pairwise preferences, addressing potential overfitting issues in DPO. 
SimPO \cite{meng2024simposimplepreferenceoptimization} and ORPO \cite{hong-etal-2024-orpo} explore simpler preference optimization objectives that do not rely on a reference model, the former simplify preference optimization by either aligning the reward function with generation likelihood, while the latter incorporates an odds ratio-based penalty to the conventional SFT loss for differentiating the generation styles between favored and disfavored responses. 
KTO \cite{ethayarajh2024_KTO} directly maximizes the utility of generations and eliminates the need for pairing preference datasets in DPO.

Despite the success of these algorithms, recent studies have highlighted the challenges posed by noisy preferences, which often arise from biased human feedback and can lead to significant performance degradation \cite{NEURIPS2022_b1efde53,wang2024secretsrlhflargelanguage}. While there has been notable progress in aligning LLMs with noisy preference data \cite{chowdhury2024provablyrobustdpoaligning,chen2024entityalignmentnoisyannotations}, existing work has not specifically addressed the alignment of LLMs with real user clicks in real-world search/recommender systems, which can be highly noisy, unbalanced and dynamic over time \cite{2022NewsRecommendersWithRoberta,10.1145/3292500.3330665,10.1145/3219819.3219823}, posing unique challenges to preference alignment.

\end{document}